\documentclass{kapedbk}
\usepackage{graphicx,amsmath,amsfonts,amssymb,bm,edbkps}
\normallatexbib
\begin{document}
\articletitle{Berezinskii-Kosterlitz-Thouless\\
              transition in 
              Josephson junction arrays}
\chaptitlerunninghead{BKT transition in JJA}

\author{Luca Capriotti}
\affil{Kavli Institute for Theoretical Physics, University of
California, Santa Barbara, CA 93106, USA}
\author{Alessandro Cuccoli and Andrea Fubini}
\affil{Dipartimento di Fisica, Universit\`a di Firenze and Unit\`a
I.N.F.M. di Firenze,\\ Via
  G. Sansone 1, 50019 Sesto Fiorentino, Italy}
\author{Valerio Tognetti}
\affil{Dipartimento di Fisica, Universit\`a di Firenze, Unit\`a
I.N.F.M. di Firenze, and Sezione I.N.F.N. di Firenze, Via G.
Sansone 1, 50019 Sesto Fiorentino, Italy}
\and 
\author{Ruggero Vaia}
\affil{Istituto di Fisica Applicata `Nello Carrara'
             del C. N. R. and  Unit\`a  I.N.F.M. di Firenze,\\
Via Madonna del Piano, 50019 Sesto Fiorentino, Italy}

\begin{abstract}
The quantum $XY$ model shows a Berezinskii-Kosterlitz-Thouless (BKT)
transition between a phase with quasi long-range order and a
disordered one, like the corresponding classical model. The effect of
the quantum fluctuations is to weaken the transition and eventually to
destroy it. However, in this respect the mechanism of disappearance of
the transition is not yet clear. In this work we address the problem
of the quenching of the BKT in the quantum $XY$ model in the region of
small temperature and high quantum coupling.  In particular, we study
the phase diagram of a 2D Josephson junction array, that is one of the
best experimental realizations of a quantum $XY$ model. A genuine BKT
transition is found up to a threshold value $g^\star$ of the quantum
coupling, beyond which no phase coherence is established. Slightly
below $g^\star$ the phase stiffness shows a reentrant behavior at
lowest temperatures, driven by strong nonlinear quantum fluctuations.
Such a reentrance is removed if the dissipation effect of shunt
resistors is included.
\end{abstract}


\section{Introduction}

\begin{figure}[h!]
\centering
\includegraphics[width=.8\textwidth]{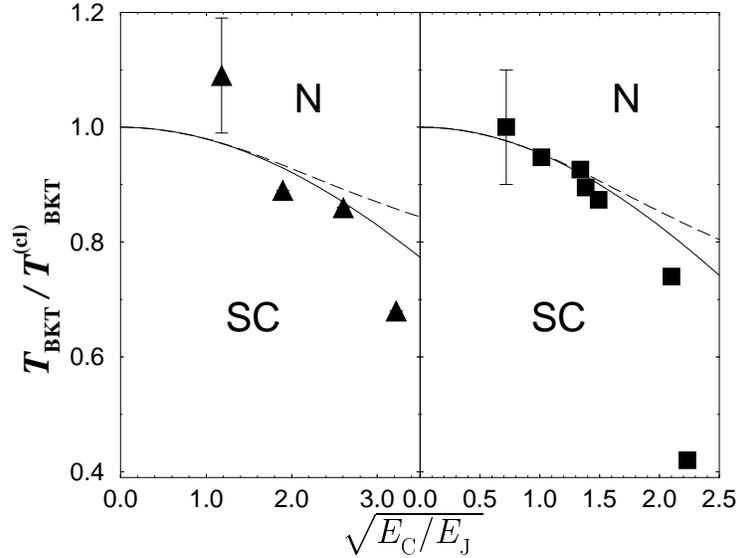}
\caption{Experimental phase diagram for a JJA on triangular (left
panel) and square (right panel) lattice \cite{fv01}. The lower
temperatures refer to the superconducting phase. The lines report
the semiclassical result of the PQSCHA~\cite{cftv00}, in the limit
\protect$R_{{\rm{S}}}\gg R_{\rm Q}=h/(2e)^2$ with $\eta=10^{-2}$
(solid), and for strong dissipation, $R_{\rm Q}/R_{{\rm{S}}}=3$
(dashed).} \label{figzero}
\end{figure}

Two-dimensional (2D) arrays of Josephson junction (JJA) are one of the
best experimental realizations of a model belonging to the $XY$
universality class and permit to check and study a variety of
phenomena related to both the thermodynamics and the dynamics of
vortices. In these systems a Berezinskii-Kosterlitz-Thouless (BKT)
transition~\cite{BKT} separates low-temperature superconducting~(SC)
state from the normal~(N) state, the latter displaying no phase
coherence~\cite{fv01}. At nanoscale size of the junctions, the quantum
fluctuations of the superconducting phases cause new interesting
features. These appear to be the consequence of the non-negligible
energy cost of charge transfer between SC islands. Indeed  
small capacitances are involved and the phase and charge are
canonically conjugate variables. A relevant effect is the progressive
reduction of the SC-N transition temperature, an example of which is
shown in Fig.~\ref{figzero}, where experimental data~\cite{zegm96} are
compared with semiclassical results~\cite{cftv00}.

Recently, fabricated arrays of nanosized junctions, both
unshunted~\cite{zegm96} and shunted~\cite{yama}, have given the
opportunity to experimentally approach the quantum (zero-temperature)
phase transition. However, the mechanism of suppression of the BKT in
the neighborhood of the quantum critical point and its connection with
the observed reentrance of the array resistance as function of the
temperature is not yet clear~\cite{fv01,zegm96,jhog89}. In this paper
we study the SC-N phase diagram by means of path-integral Monte Carlo
(PIMC)~\cite{ccftv02,ccftv04} simulations focusing the attention on the region
of strong quantum fluctuations, in order to investigate their role in
suppressing the BKT transition.

\section{The model}

The Josephson junction two-dimensional array (JJA) on
the square lattice is modeled by a quantum $XY$ model with the
following action:
\begin{equation}
 S[\bm\varphi]= \int\limits_0^{\hbar\beta}\!\! \frac{du}{\hbar}\,
 \bigg\{\!\sum_{\bm{ij}} \frac{\hbar^2C_{\bm{ij}}}{8e^2} {\dot
 \varphi_{\bm{i}}(u)\,\dot\varphi_{\bm{j}}(u)} - E_{_{\rm
 J}}\sum_{\langle{\bm{ij}}\rangle} \cos\varphi_{\bm{ij}}(u) \bigg\},
 \label{e.JJA} \end{equation} where
 $\varphi_{\bm{ij}}=\varphi_{\bm{i}}-\varphi_{\bm{j}}$ is the phase
 difference between the Josephson phases on the $\bm{i}$th and the
 $\bm{j}$th neighboring superconducting islands. The capacitance
 matrix reads
\begin{equation}
 C_{\boldsymbol{ij}}=C\,\Big[\eta\,\delta_{\boldsymbol{ij}}
 +\big( z\,\delta_{\boldsymbol{ij}}-
 {\sum}_{\boldsymbol{d}}\delta_{\boldsymbol{i},\boldsymbol{j+d}}
 \big)\Big]~,
\label{e.Cij}
\end{equation} where
$C_0\equiv\eta\,C$ and $C$ are, respectively, the self- and mutual
capacitances of the islands, and $\bm{d}$ runs over the vector
displacements of the $z\,{=}\,4$ nearest-neighbors. The standard
samples of JJA are well decribed by the limits $\eta\ll{1}$, while for
the granular films the opposite limits $\eta\gg{1}$ is more
appropriate. The quantum dynamics of the system is determined by the
Coulomb interaction between the Cooper pairs. This is described by the
kinetic term through Josephson relation
$\dot{\varphi}_{\bm{i}}=2eV_{\bm{i}}$. The Josephson coupling is
represented by the cosine term, the latter favoring Cooper-pair
tunneling across the junctions.

The quantum fluctuations are ruled by the {\em quantum coupling}
parameter $g=\sqrt{E_{_{\rm{C}}}/E_{_{\rm{J}}}}$, where
$E_{_{\rm{C}}}=(2e)^2/2C$ is the characteristic charging energy (for
$\eta\ll{1}$). In the following we use the {\em dimensionless
temperature} $t\equiv{k_{{}_{\rm B}}T/E_{_{\rm{J}}}}$. In our model
Eq.~(\ref{e.JJA}) we assume the presence of very weak Ohmic dissipation due to
the currents flowing to the substrate or through shunt
resistances~\cite{yama}, which reflects into the prescription to
consider the phase as an extended variable~\cite{fv01}.  Apart from
this, dissipative effects are negligible provided that the shunt
resistance $R_{_{\rm{S}}}\gg{R_{_{\rm{Q}}}}\,g^2/(2\pi{t})$, where
${R_{_{\rm{Q}}}}\equiv{h/(2e)^2}$ is the quantum resistance; for
smaller $R_{_{\rm{S}}}$ an explicit dissipative contribution should be
added to the action~(\ref{e.JJA}), e.g. in the form of the
Caldeira-Leggett term~\cite{fv01,cftv00},
\begin{equation}
 S _{\rm CL}[\varphi]=
 \int\limits_0^{\hbar\beta}\frac{du}{2\hbar}\int\limits_0^{\hbar\beta}\,
 du'\, \sum_{\bm{ij}}\, K_{\bm{ij}}(u-u') \,
\varphi_{\bm{i}}(u)\,\varphi_{\bm{j}}(u')\,, \label{eD.JJA}
\end{equation}
resulting in a decrease of quantum fluctuations.  The
two situations are the cases of the experiments in Ref.~\cite{zegm96}
and Ref.~\cite{yama}, respectively, where an increasing of the BKT
transition temperature was found for increasing dissipation. This can
be easily understood taking into account that the dissipative
term~(\ref{eD.JJA}) results from the coupling of the phase $\varphi_{\bm{i}}$
with environmental variables (the degrees of
freedom of the dissipative bath), constituting an implicit 
measurement of $\varphi_{\bm{i}}$ .

\section{Numerical simulations}

\begin{figure}[t]
\centering
\includegraphics[width=.8\textwidth]{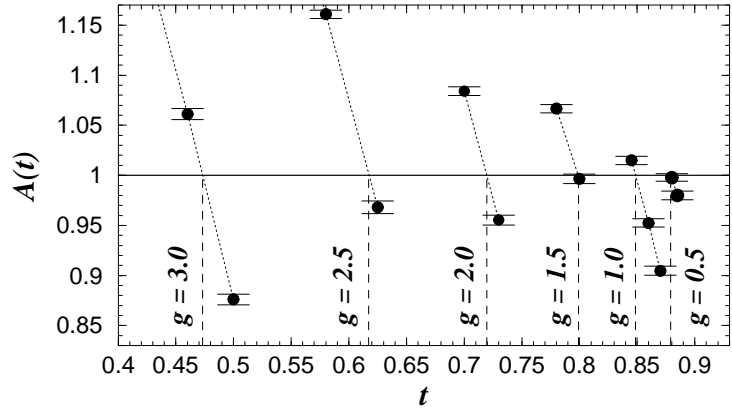}
\caption{Fitting parameter $A(t)$ for different values of $g$
ranging from $0.5$ to $3.0$. The BKT transition temperature is
obtained by the condition $A(t_{{}_{\rm BKT}})\,{=}\,1$.
\label{figtwoa}}
\end{figure}
The numerical data for the BKT transition temperature are obtained
using PIMC simulations on $L\times L$ lattices (up to $L=96$) with
periodic boundary conditions. Thermodynamic averages are obtained by
MC sampling of the partition function after discretization of the
Euclidean time $u\in[0,\beta\hbar]$ in $P$ slices $\,\hbar\beta/P\,$,
where $P$ is the Trotter number. However, the actual sampling is made
on imaginary-time Fourier transformed variables on a lattice using the
algorithm developed in Ref.~\cite{ccftv02}. The move amplitudes are
independently chosen and dynamically adjusted for each Fourier
component; this procedure turns out to be very efficient to reproduce
the strong quantum fluctuations of the paths in the region of high
quantum coupling \,$g$. Indeed, test simulations with the standard
PIMC algorithm showed serious problems of ergodicity, though
eventually giving the same results. The approach through Fourier
PIMC becomes more and more suitable and effective when
dissipation is inserted, because the non-local action~(\ref{eD.JJA})
becomes local in Fourier space. Additional details about the numerical
method are given in the Appendix. An over-relaxation
algorithm~\cite{over} over the zero-frequency mode has also been
implemented in order to effectively reduce the autocorrelation times.

A very sensitive method to determine the transition temperature is
provided by the scaling law of the helicity modulus $\Upsilon$
(a quantity proportional to the phase stiffness),
\begin{equation}
 \Upsilon = \frac{1}{E_{_{\rm J}}}
\left(\frac{\partial^2 F}{\partial k_0^2}\right)_{k_0=0}~,
\label{e.Y}
\end{equation}
which measures the response of the free energy $F$ (per unit volume)
when a uniform twist $k_0$ along a fixed direction $\bm{u}$ is applied
to the boundary conditions (i.e.,
$\varphi_{\bm{i}}\to\varphi_{\bm{i}}+k_0\bm{u}{\cdot}\bm{i}$, with the
unitary vector $\bm{u}$).

\begin{figure}[b]
\centering
\includegraphics[width=.8\textwidth]{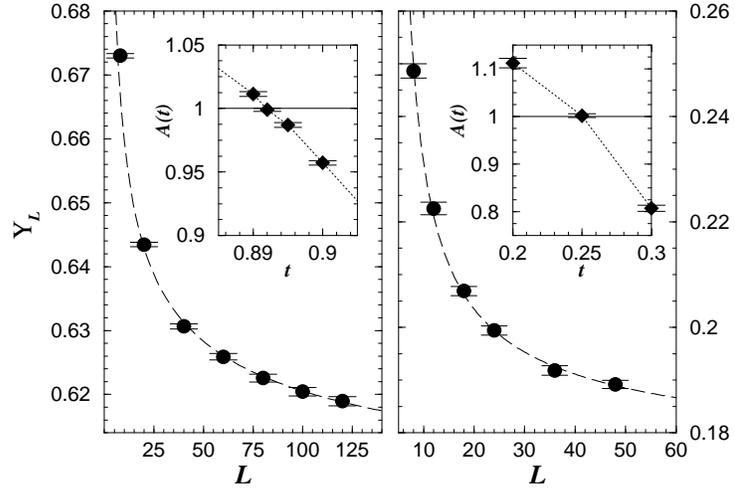}
\caption{
Size scaling of the helicity modulus $\Upsilon_L$ at the transition
temperature. Symbols are PIMC data and the dashed-lines are the
one-parameter fit with Eq.~\eqref{e.YL}, i.e. with $A(t)=1$. Left
panel: $g=0$ and $t=0.892$ [$L_0=0.456(6)$]; right panel: $g=3.4$ and
$t=0.25$ [$L_0=3.32(3)$]. The insets show $A(t)$ for different
temperatures, using the two-parameter fit (see text).
\label{figtwo}}
\end{figure}
\begin{figure}[t]
\centering\vskip.2in
\includegraphics[width=.65\textwidth]{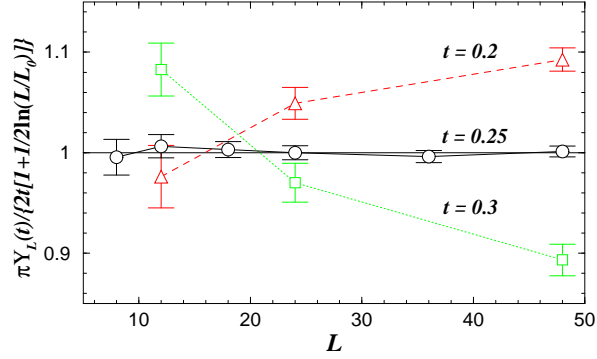}
\caption{
 Helicity modulus $\Upsilon_L$ divided by the best fit with the
 expression~(\ref{e.YL}) for $g\,{=}\,3.4$ and different
 temperatures: $\triangle$, $\bigcirc$, and $\square$ correspond to
 $t=0.2$, $0.25$, $0.3$, respectively.
\label{figthree}}
\end{figure}

The PIMC estimator for $\Upsilon$ is easily obtained, in analogy to
that of Ref.~\cite{RJ96}, by derivation of the path-integral
expression of the partition function (see Appendix). Kosterlitz's
renormalization group equations provide the critical scaling law for
the finite-size helicity modulus $\Upsilon_L$:
\begin{equation}
 \frac{\Upsilon_L(t_{{}_{\rm BKT}})}{t_{{}_{\rm BKT}}} =
 \frac{2}{\pi}\left(1+\frac{1}{2\ln(L/L_0)}\right)~,
\label{e.YL}
\end{equation}
where $L_0$ is a non-universal constant. Following Ref.~\cite{harada},
the critical temperature can be found by fitting $\Upsilon_L(t)/t$ vs
$L$ for several temperatures according to Eq.~\eqref{e.YL} with a
further multiplicative fitting parameter $A(t)$. In this way, the
critical point can be determined by searching the temperature such
that $A(t_{{}_{\rm BKT}})\,{=}\,1$, as illustrated in
Figs~\ref{figtwoa} and~\ref{figtwo}. 
Using this procedure the critical temperature can be determined
with excellent precision. For instance in the classical case we
get $t_{{}_{\rm BKT}}(g{=}0)=0.892(2)$, in very good agreement
with the most accurate results from classical
simulations~\cite{gupta}. Also in the regime of strong quantum
coupling, $g=3.4$, the PIMC data for
$\Upsilon_L(t_{_{\rm{BKT}}}\,{=}\,0.25)$ are very well fitted by
Eq.~\eqref{e.YL}, as shown in Fig.~\ref{figthree}. Moreover, this
figure points out the sensitivity of this method to identify
$t_{_{\rm{BKT}}}$: at temperature higher (lower) than the critical
one the helicity modulus decreases (increases) much faster with
$L$ than $\Upsilon_L(t_{_{\rm{BKT}}})$. At higher values of the
quantum coupling, $g>g^\star$, the helicity modulus scales to zero
with $L\to\infty$ and $P\to\infty$ at any temperature~\cite{ccftv04}.

\begin{figure}[ht]
\centering\vskip.2in
\includegraphics[width=.8\textwidth]{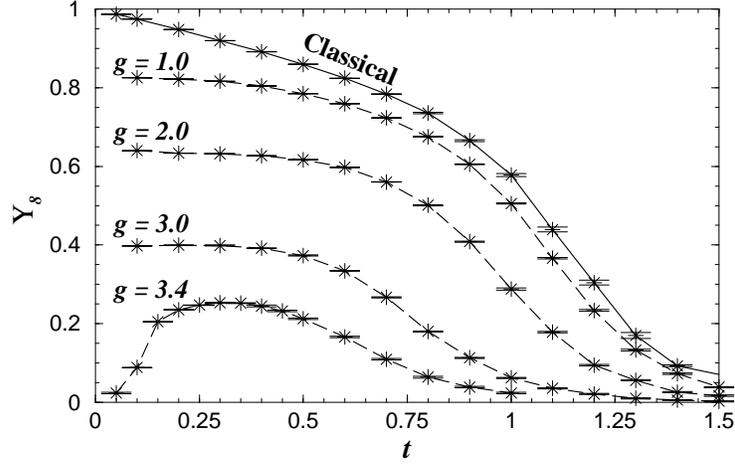}
\caption{
 Temperature behavior of the helicity modulus $\Upsilon_8(t)$ on a
 $8\,{\times}\,8$ lattice, for different values of $g$. The data
 are results from the Trotter extrapolation.
\label{figfour}}
\end{figure}

\section{Results}

We have found significant differences with the standard BKT
theory. In the regime of strong quantum fluctuations, a range of
coupling values, $3.2\lesssim{g}\lesssim{g^\star}$, is found in
which the helicity modulus displays a non-monotonic temperature
behavior.

In Fig.~\ref{figfour}, $\Upsilon_L(t)$ is plotted for different
values of $g$ on the $8\times8$ cluster: up to $g=3.0$ it shows a
monotonic behavior similar to the classical case, where thermal
fluctuations drive the suppression of the phase stiffness. In
contrast, for $g\,{=}\,3.4$, the helicity modulus is suppressed at
low temperature, then it increases up to $t\sim0.2$; for further
increasing temperature it recovers the classical-like behavior and a
standard BKT transition can still be located at $t\sim0.25$
(Figs.~\ref{figtwo} and~\ref{figthree}). A reentrance of the phase
stiffness was already found for a related model in Ref.~\cite{RJ96},
but the authors concluded that the drop of the helicity modulus at
lowest temperatures was probably due to the finiteness of the
Trotter number $\,P\,$.

\begin{figure}[ht]
\centering
{\includegraphics[width=.7\textwidth]{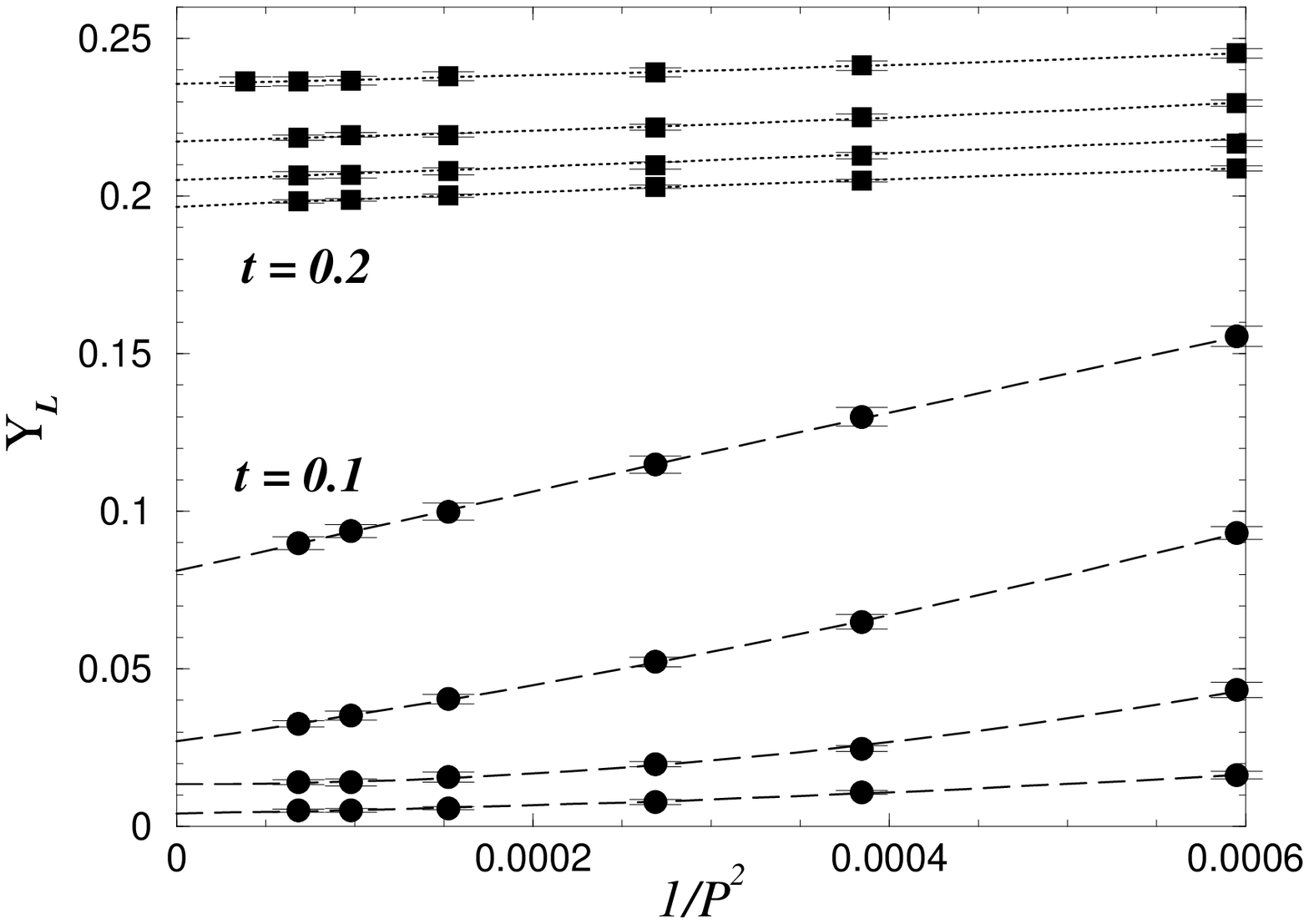}
\caption{
 Trotter-number extrapolation of $\Upsilon_L$ for $g=3.4$. Two
 series of data for $t=0.1$ ({\Large $\bullet$}) and $0.2$
 ($\blacksquare$) are reported, for four different lattice sizes:
 from the top to the bottom $L=8,10,12,14$. The lines are weighted
 quadratic fits.}
\label{figfive}}
{\includegraphics[width=.7\textwidth]{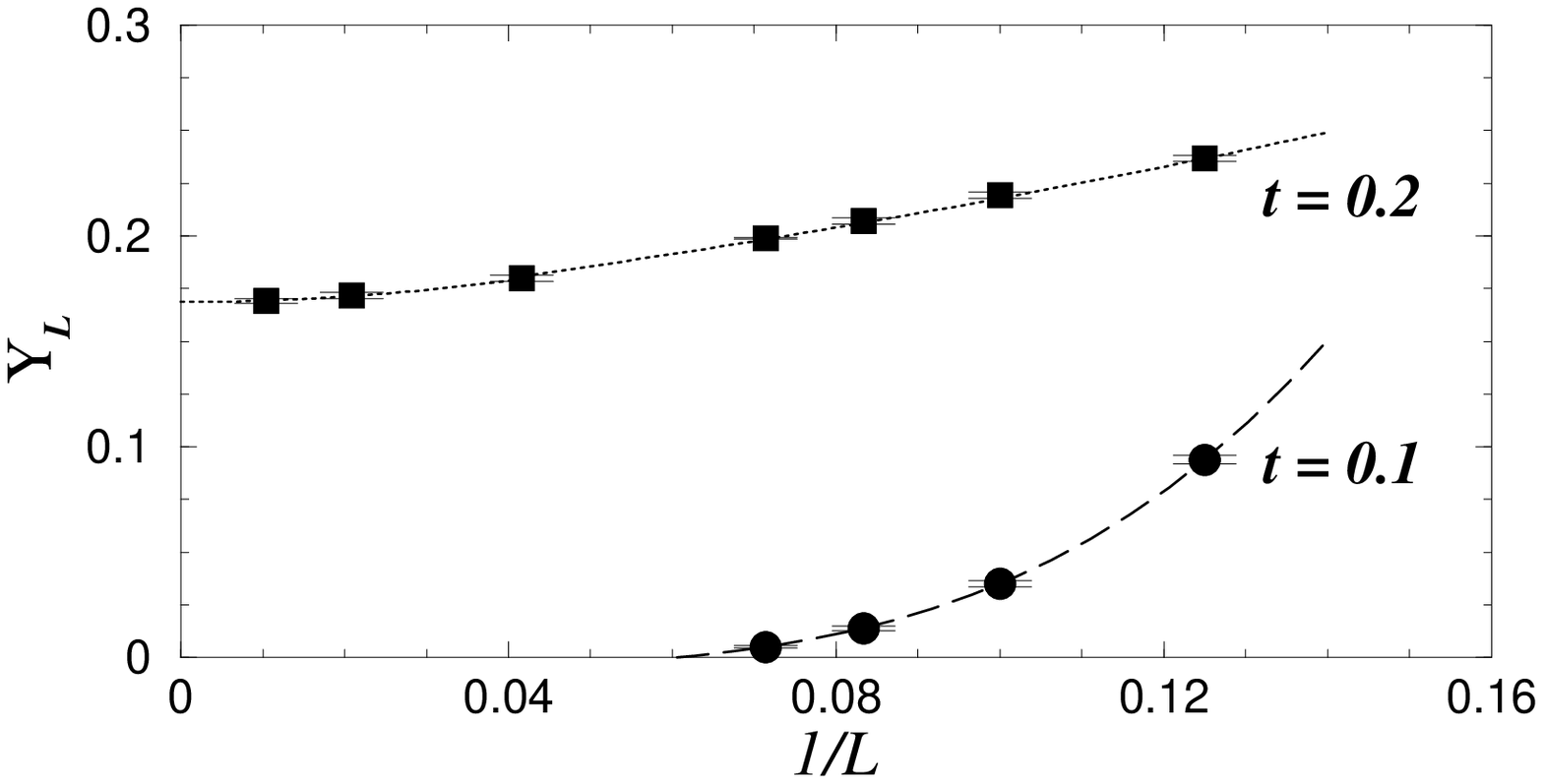}
\caption{
Finite-size scaling of the helicity modulus $\Upsilon_L$ for
$g=3.4$ at fixed $P=101$. The lines are guides for the eye.}
\label{figsix}}
\end{figure}

Systematic extrapolations in the Trotter number and in the lattice
size have been done, and presented in Figs.~\ref{figfive}
and~\ref{figsix} for $g=3.4\,$, in order to ascertain this point. In
particular, we did not find any anomaly in the finite-$P$ behavior:
the extrapolations in the Trotter number appear to be well-behaved,
in the expected asymptotic regime $\,{\cal
O}(1/P^2)\,$~\cite{trotta}, for $P\gtrsim 60$ (Fig.~\ref{figfive}).
Moreover, the extrapolation to infinite lattice-size shown in
Fig.~\ref{figsix} clearly indicates that $\Upsilon_L$ scales to zero
at $t=0.1$, while it remains finite and {\em sizeable} at $t=0.2$.
therefore, the outcome of our analysis is opposite to that of
Ref.~\cite{RJ96}, i.e., we conclude that the reentrant behavior of
the helicity modulus appears to be a genuine effect present in the
model, rather than a finite-Trotter or finite-size artifact.

\begin{figure}[ht]
\centering\vskip.2in
\includegraphics[width=.7\textwidth]{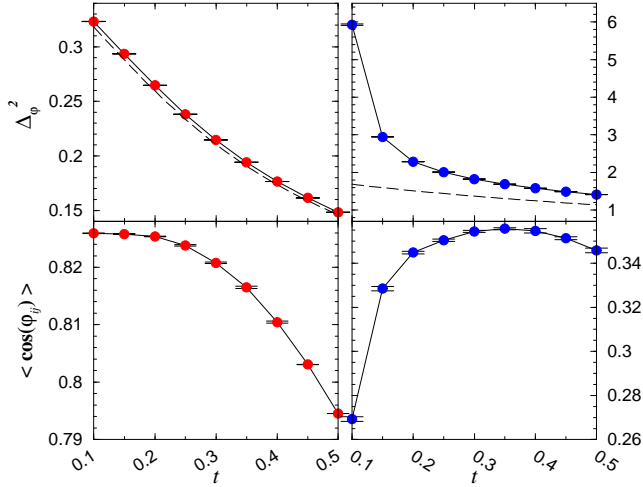}
\caption{
Top panels: $\Delta_\varphi^2$ vs $t$; bottom panels:
$\langle\cos\,\varphi_{\bm{ij}}(u)\rangle$ vs $t$. The quantum
coupling is $g=1.0$ in the left panels and $g=3.4$ in the right
panels. The circles are PIMC data and the dashed lines are PQSCHA
results.
\label{figseven}}
\end{figure}

In order to understand the physical reasons of the reentrance
observed in the phase stiffness, we have studied the following two
quantities:
\begin{eqnarray}
{\cal A}  &=&
\big\langle\,\cos\,\varphi_{\bm{ij}}(u)\,\big\rangle~,
\label{e.cos}
\\
  \Delta_\varphi^2&=& \big\langle\,
 (\varphi_{\bm{ij}}(u)-\bar{\varphi}_{\bm{ij}})^2\,\big\rangle~,
\label{e.D}
\end{eqnarray}
with
\begin{equation}
\bar{\varphi}_{\bm{ij}}=\frac 1{\hbar\beta}
\int_0^{\hbar\beta}\!du~\varphi_{\bm{ij}}(u)\,
\label{e.ZFC}
\end{equation}
and $\bm{ij}\,$ nearest-neighbor sites. The first quantity
$\,{\cal A}\,$, is a measure of the total (thermal plus quantum)
short-range fluctuations of the Josephson phase and its maximum
occurs where the overall fluctuations are lowest. The second
quantity represents instead the pure-quantum spread of the phase
difference between two neighboring islands and has been recently
studied in the single junction problem~\cite{hz02}; more
precisely, $\Delta_\varphi^2\,$ measures the fluctuations around
the static value (i.e., the zero-frequency component of the
Euclidean path), it is maximum at $t\,{=}\,0$ and tends to zero in
the classical limit, i.e., $(g/t)\to{0}$.
\begin{figure}[hb]
\centering
\includegraphics[width=.5\textwidth]{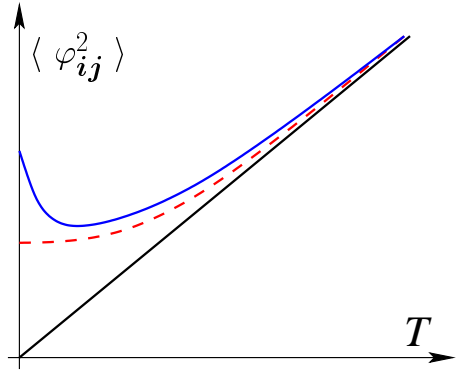}
\caption{Qualitative picture of the behavior of the total mean
square fluctuation of the nearest-neighbor phase difference. The
straight line refer to the classical fluctuations. The lower
(dashed) curve refers to the low-coupling case. The upper curve
represents  strongest couplings.
$\langle\varphi_{\bm{ij}}^2\rangle$. \label{qualitative}}
\end{figure}

The quantities~(\ref{e.cos}) and~(\ref{e.D}) on a $8\,{\times}\,8$
lattice are compared in Fig.~\ref{figseven} for two values of the
quantum coupling, in the semiclassical ($g=1.0$) and in the extreme
quantum ($g=3.4$) regime. In the first case
$\,{\cal{A}}=\langle\cos\varphi_{\bm{ij}}(u)\rangle\,$ decreases
monotonically by increasing $t$ and the pure-quantum phase spread
$\Delta_\varphi^2$ shows a semiclassical linear behavior which is
correctly described by the PQSCHA. At variance with this, at $g=3.4$,
where the reentrance of $\Upsilon(t)$ is observed, $\,{\cal A}=
\langle\cos\varphi_{\bm{ij}}(u)\rangle\,$ shows a pronounced maximum
at finite temperature. Besides the qualitative agreement with the
mean-field prediction of Ref.~\cite{fms84}, we find a much stronger
enhancement of the maximum above the $t\,{=}\,0$ value. This
remarkable finite-$t$ effect can be explained by looking at
$\Delta_\varphi^2(g{=}3.4)$ (Fig.~\ref{figseven}, top-right panel).
In fact its value is an order of magnitude higher than the one in the
semiclassical approximation and, notably, it is strongly suppressed by
temperature in a qualitatively different way from
$\Delta_\varphi^2(g{=}1.0)$: i.e., the pure-quantum contribution to
the phase fluctuations measured by $\Delta_\varphi^2$ decreases much
faster than the linearly rising classical (thermal) one resulting in a
global minimum of the total fluctuations. The qualitative behavior of
the total mean square fluctuation of the nearest-neighbor phase
difference vs. temperature is sketched in Fig.$\,$\ref{qualitative}. This
single-junction effect, in a definite interval of the quantum coupling
($3.2\lesssim{g}\lesssim3.4$), is so effective to drive the reentrance
of the phase stiffness. As for the transition in the region of 
high quantum fluctuations and low temperature, the
open symbols in Figs.~\ref{figone} and~\ref{figlast} represent the
approximate location of the points $(t,g)$ where $\Upsilon(t)$ becomes
zero within the error bars: in their neighborhood we did not find any
BKT-like scaling law. This fact opens two possible interpretations:
($i$) the transition does not belong to the $XY$ universality class;
($ii$) it does, and in this case the control parameter is not the
(renormalized) temperature, but a more involute function of both $t$
and $g$.

\section{The phase diagram}

\begin{figure}[ht]
\centering
\includegraphics[width=.9\textwidth]{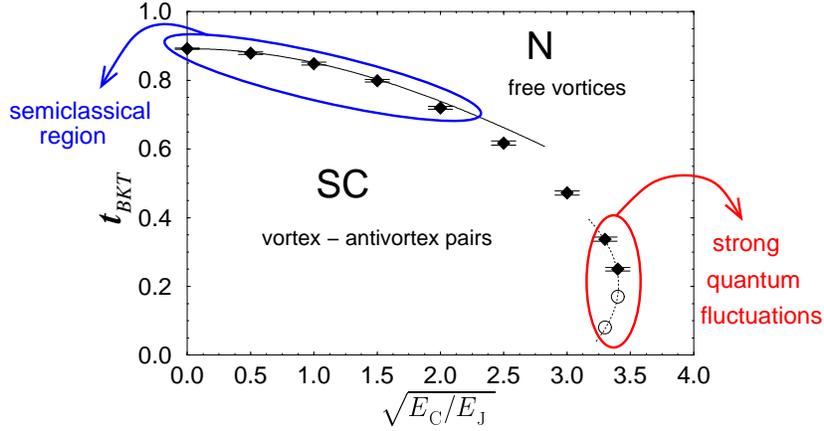}
\caption{Phase diagram for a  square lattice Josephson junction
arrays  in the limit \protect$R_{{\rm{S}}}\gg R_{\rm Q}=h/(2e)^2$
with $\eta=10^{-2}$ . The symbols are our PIMC results and the
line reports the semiclassical result of the
PQSCHA~\cite{cftv00}.} \label{figone}
\end{figure}

\begin{figure}[ht]
\centering
\includegraphics[width=.7\textwidth]{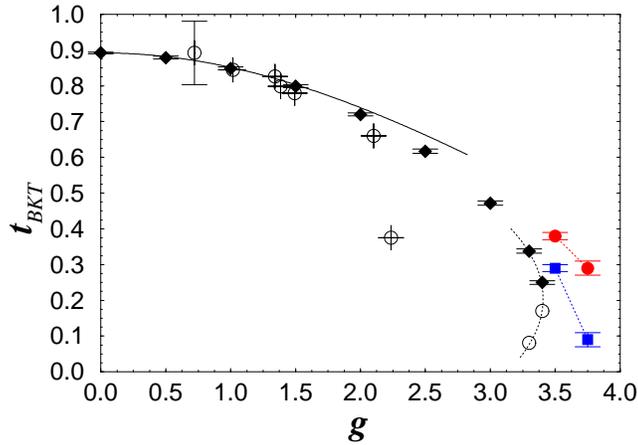}
\caption{Phase diagram for a  square lattice Josephson junction
arrays. Crossed circles are the experimental data~\cite{fv01}. The
other filled symbols are our PIMC results, in the limit
$R_{{\rm{S}}}\gg{R_{\rm{Q}}}=h/(2e)^2$ (diamonds), for $R_{\rm
Q}/R_{{\rm{S}}}=0.25$ (squares) and $R_{\rm Q}/R_{{\rm{S}}}=0.5$
(circles), all with $\eta=10^{-2}$. The solid line reports the
semiclassical non dissipative result of the PQSCHA~\cite{cftv00}. }
\label{figlast}
\end{figure}

Let us now describe our resulting phase diagram~\cite{ccftv04}, displayed in Fig.~\ref{figone} together with the semiclassical results valid at low coupling~\cite{cftv00} and in Fig.~\ref{figlast} together with the experimental data~\cite{zegm96} and our first PIMC outcomes including the dissipation effect.

At high temperature, the system is in the normal state with
vanishing phase stiffness and exponentially decaying phase
correlations, $\langle\varphi_{\bf{i}}\varphi_{\bf{j}}\rangle$.
By lowering $t$, for $g\le{g^\star}\,{\simeq}\,3.4$, the system
undergoes a BKT phase transition at $t_{_{\rm{BKT}}}(g)$ to a
superconducting state with finite stiffness and power-law
decaying phase correlations. When $g$ is small enough
(semiclassical regime), the critical temperature smoothly
decreases by increasing $g$ and it is in remarkable agreement with
the predictions of the pure-quantum self-consistent harmonic
approximation (PQSCHA)~\cite{cftv00}. For larger $g$ (but still
$g<g^\star$) the semiclassical treatment becomes less accurate and
the curve $t_{{}_{\rm BKT}}(g)$ shows a steeper reduction, but the
SC-N transition still obeys the standard BKT scaling behavior.
Finally, for $g>g^\star$ a strong quantum coupling regime with no
sign of a SC phase is found. Surprisingly, the BKT critical
temperature does not scale down to zero by increasing $g$ (i.e.,
$t_{{}_{\rm{BKT}}}(g^\star)\neq{0}$): by reducing the temperature
in the region $3.2\lesssim{g}\lesssim{g^\star}$, phase coherence
is first established, as a result of the quenching of thermal
fluctuations, and then destroyed again due to a dramatic
enhancement of quantum fluctuations near $t=0$. This is evidenced
by a reentrant behavior of the stiffness of the system, which
vanishes at low and high $t$ and it is finite at intermediate
temperatures. The open symbols in Fig.~\ref{figone} mark the
transition between the finite and zero stiffness region when $t$
is lowered.

When the interaction with a heat bath given by Eq.$\,$ (\ref{eD.JJA})
is present through a variable shunt resistance, the quantum phase
fluctuations are decreased by the dissipation so that the BKT
transition temperature rises. This is well reproduced by PQSCHA
approach~\cite{cftv00} and turns out to be in qualitative agreement
with recent experiments~\cite{yama}.  As shown in Fig.~\ref{figlast},
the presence of intermediate ($R_{\rm Q}/R_{\rm S}=0.25$) and high
($R_{\rm Q}/R_{\rm S}=0.5$) dissipation causes the disappearance of
the reentrance, in agreement with the experiments of Takahide {\em et
al.}~\cite{yama}. Their data do not show any anomalous reentrant
behavior like those ones of the experiments on {\em unshunted} samples
in Ref.~\cite{zegm96}.  Further investigations are needed to answer
the question about the nature of the transition in the reentrance zone
and to determine at which values of the dissipation the reentrant
behavior is washed out.

\section{Summary}

In summary, we have studied a model for a JJA in the
quantum-fluctuation dominated regime by means of Fourier
path-integral Monte Carlo simulations. The BKT phase transition has
been followed increasing the quantum coupling $g$ up to a critical
value $g^\star\,{\sim}\,3.4$ where $t_{{}_{\rm BKT}}\sim0.25$; above
$g^\star$ no traces of BKT critical behavior have been observed.
Remarkably, in the regime of strong quantum coupling
($3.2\lesssim{g}\lesssim3.4$) phase coherence is established only in
a finite range of temperatures, disappearing at higher $T$, with a
genuine BKT transition to the normal state, and at lower $T$, due to
a nonlinear quantum mechanism. This effect is destroyed
by the presence of a relevant dissipation.

\begin{acknowledgments}
The authors acknowledge thorough discussions and useful suggestions
from G.~Falci, R.~Fazio, M.~M\"user, T.~Roscilde, and U.~Weiss. We
thank H.~Baur and J.~Wernz for assistance in using the MOSIX cluster
in Stuttgart. L.C. was supported by NSF under Grant No. DMR02-11166.
This work was supported by the MIUR-COFIN2002 program.
\end{acknowledgments}

\chapappendix{PIMC in the Fourier space for JJA}

In this appendix we derive the discretized expressions of the
observables to be measured by means of the simulations based on the
Metropolis algorithm. The first step is to discretize the Euclidean
time in $P$ slices, $u_l=(\hbar\beta/P)l$, and to perform a discrete
Fourier transform
\begin{equation}
\varphi_{{\bm i},\alpha}=\frac1P \sum_{l=1}^P \varphi_{{\bm i},l}\, {\rm
e}^{i\frac{2\pi}P l\alpha}~,
\label{e.DFT}
\end{equation}
where $\varphi_{{\bm i},l}=\varphi_{\bm i}(u_l)$, and $\varphi_{{\bm
i},\alpha}$ satisfies the conditions of {\em periodicity}, i.e. $\varphi_{{\bm
i},\alpha}=\varphi_{{\bm i},\alpha+P}$, and {\em reality} of $\varphi_{{\bm
i},l}$, i.e. $\varphi_{{\bm i},\alpha}=\varphi_{{\bm
i},-\alpha}^*$. Thus the discretized path can be written as a sum over
different frequency components
\begin{equation}
\varphi_{{\bm i},l} = \bar{\varphi}_{{\bm i}} + 2
\sum_{\alpha=1}^{N} \Re{\rm e}\left[ \varphi_{{\bm i},\alpha} {\rm
e}^{-i\frac{2\pi}P l\alpha} \right] =  \bar{\varphi}_{{\bm i}} + 2
\sum_{\alpha=1}^{N} \left[ a_{{\bm i},\alpha}
\cos\frac{2\pi l\alpha}P + b_{{\bm i},\alpha}
\sin\frac{2\pi l\alpha}P \right]~,
\label{e.IDFT}
\end{equation}
where $\bar{\varphi}_{{\bm i}}$ is the zero-frequency component of the
Euclidean path Eq.~(\ref{e.ZFC}), and we choose an odd Trotter number
$P=2N+1$. The advantage of using expression~(\ref{e.IDFT}) is
twofold. On the one hand the dissipative term of the action~(\ref{eD.JJA})
that is non-local in time becomes diagonal. On the other hand the
sampling can be performed independently on each frequency component,
dynamically adjusting each move amplitude. This method make the
sampling very effective especially in the region of strong quantum
fluctuations, where the
main contribution to the path comes from $\{a_{{\bm i},\alpha}\}$ and
$\{b_{{\bm i},\alpha}\}$.

Using expression~(\ref{e.IDFT}), the JJA action~(\ref{e.JJA}) plus the
dissipative term~(\ref{eD.JJA}) reads
\begin{equation}
S[{\bm \varphi}] = \sum_{\bm{ij}}\sum_{\alpha=1}^N T_{\bm{ij},\alpha}
(a_{{\bm i},\alpha}a_{{\bm j},\alpha}+b_{{\bm i},\alpha}b_{{\bm
j},\alpha}) + \frac{\beta E_{_{\rm J}}}{P}
\sum_{\langle\bm{ij}\rangle} \sum_{l=1}^P
\big(1-\cos\varphi_{\bm{ij},l}\big)~,
\label{e.DS}
\end{equation}
where $\varphi_{\bm{ij},l} = \varphi_{{\bm i},l}{-}\varphi_{{\bm
j},l}$, and the ``kinetic'' matrix is
\begin{equation}
T_{\bm{ij},\alpha} = \frac{P^2}{\beta e^2}C_{\boldsymbol{ij}}
\sin^2 \frac{\pi \alpha}{P} + \beta K_{\bm{ij},\alpha}~,
\label{e.Tij}
\end{equation}
and $K_{\bm{ij},\alpha}$ is the discrete FT of the dissipative kernel
matrix $K_{\bm{ij}}(u{-}u')$. All the macroscopic thermodynamic
quantities are obtained through the estimators generated from the
discretized action~(\ref{e.DS}). For instance the estimator of the
helicity modulus $\Upsilon$ is obtained applying the
definition~(\ref{e.Y}) to the discretized free energy per unit volume
\begin{equation}
F = \frac1\beta \ln{\cal Z} = \frac1\beta \ln\left[C \int {\cal D}{\bm
\varphi}\; {\rm e}^{-S[{\bm \varphi}]} \right]~,
\label{e.F}
\end{equation}
with the normalization constant $C$. Eventually we get
\begin{align}
\Upsilon_P & = t \Big(\frac1{\cal Z}\frac{\partial {\cal Z}}{\partial
k_0}\Big\arrowvert_{k_0=0}\Big)^2 - t \frac1{\cal Z}
\frac{\partial^2 {\cal Z}}{\partial k_0^2}\Big\arrowvert_{k_0=0}
\notag \\
&= \frac1P \sum_{\langle\bm{ij}\rangle}\sum_{l=1}^P 
\cos\varphi_{\bm{ij},l} - \frac1{tP^2}
\Big(\sum_{\langle\bm{ij}\rangle}\sum_{l=1}^P
\sin\varphi_{\bm{ij},l}\Big)^2~.
\label{e.DY}
\end{align}

\end{document}